\documentclass[twocolumn,showpacs,preprintnumbers,amsmath,amssymb]{revtex4}
\usepackage{dcolumn}% Align table columns on decimal point
\usepackage{bm}% bold math
\usepackage{latexsym}
\usepackage{amsfonts}
\usepackage[dvips]{graphicx}
\usepackage[usenames]{color}
\usepackage{makeidx}
\newcommand{\pa}{\partial}

\newcommand{\ket}{\rangle }
\newcommand{\bra}{\langle }

\newcommand{\ve}{\varepsilon}

\newcommand{\up}{\uparrow}
\newcommand{\dw}{\downarrow}

\newcommand{\Vect}[1]{\mbox{\boldmath$#1$}}
\newcommand{\ihbar}{\frac{i}{\hbar}}

\begin{document}

%\preprint{APS/123-QED}

\title{Ground-State Decay Rate for the Zener Breakdown in  
Band and Mott Insulators}
\author{Takashi Oka$^*$ and Hideo Aoki}
\address{Department of Physics, University of Tokyo, Hongo, Tokyo 113-0033, 
Japan}

\date{\today}
\begin{abstract}
\noindent 
Non-linear transport of electrons in strong electric 
fields, as typified by dielectric breakdown, 
is re-formulated in terms of the ground-state decay rate
originally studied by Schwinger in non-linear QED. 
We discuss the effect of electron interaction on
Zener tunneling by 
comparing the dielectric breakdown of the band insulator
and the Mott insulator, where the latter is studied by
the time-dependent density-matrix renormalization group.
The relation with the Berry's phase theory of 
polarization is also established.

\end{abstract}

\maketitle

\maketitle
{\it Introduction ---} 
While there is a mounting body of interests in non-linear responses in 
many-body systems, the 
dielectric breakdown of Mott insulators is conceptually 
interesting in a number of ways.  For band insulators 
the dielectric breakdown has been well understood in terms of 
the Zener tunneling\cite{Zener1934} 
across the valence and conduction bands, which triggers 
the breakdown.  By contrast, 
non-linear properties in {\it strongly correlated 
electron systems} should be qualitatively different,  
since the excitation gap 
in the insulating side of Mott's metal-insulator transition, 
arising from the electron-electron repulsion, 
is something totally different from the 
band gap in its origin and nature.  Experimentally, 
Taguchi {\it et al.} observed a dielectric breakdown 
by applying strong electric fields to typical Mott insulators 
(quasi-one-dimensional cuprates, Sr$_2$CuO$_3$, SrCuO$_2$)\cite{tag}.
The present authors with Arita proposed theoretically that the 
phenomenon may be explained in terms of a Zener 
tunneling for many-body levels, and obtained the 
threshold field strength \cite{Oka2003}.  

Now, there is an important difference between the Zener breakdowns 
in the band and Mott insulators: 
excitations (i.e., electrons and holes produced by the field) 
move freely in the former, while they must interact 
with surrounding electrons in the latter and become dissipated. 
In other words, while we have only to worry about the 
valence-conduction gap, we have to deal with many charge 
gaps among the many-body levels when the interaction is 
present.  
Several authors, including the 
present authors, have shown, with effective models, that 
a suppression of quantum tunneling due 
to the quantum interference should occur in many-level systems driven 
by external forces 
(see \cite{Gefan1987,Cohen2000,Oka2004a} and refs. therein).

The tunneling rate, first obtained by Zener, 
gives the amount of the excitations, 
and is proportional to the leakage current 
if all the excitations are absorbed by electrodes\cite{Zener1934}. 
So the rate is a crucial quantity, but 
theoretical studies have been quite scant: 
Even for band insulators, the quantity is only calculated for 
systems having a simple band dispersion. 
More importantly, 
Zener's tunneling rate was based on a one-body
WKB approach, so we must extend the formalism to many-body systems. 
Specifically, the actual breakdown should be related to the 
scattering between excited states in many-body systems as noted above.  
While this is shown to lead to a suppression of the current 
in a toy (quantum-walk) model\cite{Oka2004a}, we are still badly 
in need of studies for {\it microscopic} models, since the 
existing calculation for the Hubbard model\cite{Oka2003} was limited to small 
systems (hence to short-time behaviors), while 
the long-time behavior, affected strongly by scattering 
between excited states, is in fact relevant. 

In this Letter, we propose to determine the 
tunneling rate from the {\it ground-state-to-ground-state 
transition amplitude}, whose long-time asymptotic defines the 
{\it effective Lagrangian}. 
The effective-Lagrangian approach was 
evoked by Heisenberg and Euler in their 
study of quantum electrodynamics (QED) in
strong electric fields\cite{Heisenberg1936}. 
This was extended by Schwinger to obtain the decay rate of the QED 
vacuum\cite{Schwinger1951}. It is well known that Schwinger's 
electron-positron creation rate can be understood by 
the Landau-Zener mechanism (see e.g. 
\cite{Rau1996}). 
What we have done here is the following: 
(i) We first express the effective Lagrangian for band insulators, 
which contains higher-order terms in the Landau-Zener's 
tunneling probability. 
Since breakdown of band insulators is a well understood
subject, the main aim of this section is to introduce the notations
for later discussions.
(ii) We then move on to a Mott insulator 
in strong electric fields.  The ground-state decay rate 
is obtained for a microscopic (one-dimensional half-filled Hubbard) model 
with the time-dependent density-matrix renormalization group\cite{Vidal2004,White2004}. 
From this we determine the threshold electric 
field to construct the ``dielectric breakdown phase diagram".  
(iii) Finally  we comment on an intriguing 
link between Heisenberg-Euler-Schwinger's 
effective Lagrangian approach and a recent, Berry's phase 
approach to polarization proposed in 
Refs.\cite{Resta1992,KingSmith1993,Resta1998,Resta1999,Nakamura2002}.

%%%%%%%%%%%%%%%%%%%%%%%%%%%%%%%%%%%%%%%%%%%%%%%%%%
{\it Dielectric Breakdown of a Band Insulator ---}
%%%%%%%%%%%%%%%%%%%%%%%%%%%%%%%%%%%%%%%%%%%%%%%%%%
We start with the dielectric breakdown of 
band insulators in an electric field $\Vect{F}$
within the effective-mass approximation. 
For simplicity we take a pair of hyperbolic bands 
$\ve_\pm(\Vect{k})=\pm \sqrt{V^2+v^2k^2}$ 
(considered here in $d$ spatial dimensions), 
where $V$ is the band gap, $-(+)$ represents the 
valence (conduction) band, 
and $v$ the asymptotic slope of the dispersion.  

We first obtain the ground-state-to-ground-state 
transition amplitude with the time-dependent gauge
in the periodic boundary condition. 
There, a time-dependent AB-flux measured by the 
flux quantum, $\phi(\tau)=FL\tau/h$ 
(with the electronic charge put equal to $e=1$,
$L$ the system size) is introduced to 
induce an electric field $F$, which makes the Hamiltonian 
time dependent as 
$
H(\phi(\tau))=
\sum_{\bf{k},\alpha=\pm}\ve_\alpha(\Vect{k}+\frac{2\pi}{L}\phi(\tau)\Vect{e}_{\parallel})
c_{\alpha}^\dagger(\Vect{k})c_{\alpha}(\Vect{k}).
$
Here $\Vect{e}_{\parallel}$ is the 
unit vector parallel to $\Vect{F}$, 
and $c_{\alpha}^\dagger(\Vect{k})$ the creation operator with 
spin indices dropped. 
If we denote the ground state of $H(\phi)$ as $|0;\phi\ket$ 
and its energy as $E_0(\phi)$, the ground-state-to-ground-state 
transition amplitude is defined as
\begin{eqnarray}
\Xi (\tau)
=\bra 0;\phi(\tau)|\hat{T}e^{-\frac{i}{\hbar}
\int_0^\tau H(\phi(s))ds}|0;\phi(0)\ket e^{\frac{i}{\hbar}\int_0^\tau E_0(\phi(s))ds},
\label{ggamplitude}
\end{eqnarray}
where $\hat{T}$ stands for the time ordering.
The effective Lagrangian $\mathcal{L}(F)$ for 
the quantum dynamics is defined from the asymptotic behavior, 
$\Xi (\tau)\sim e^{\ihbar \tau L^d\mathcal{L}(F)}$.  
The imaginary part of the effective Lagrangian gives
the decay rate $\Gamma(F)/L^d=2\mbox{Im}\;\mathcal{L}(F)/\hbar$ 
\`{a} la Callan-Coleman\cite{CallanColeman1977} 
(see also \cite{Niu1998}).
$\Gamma(F)/L^d$ gives the rate of the exponential decay 
for an unstable vacuum (ground state), which, in the case of Zener 
tunneling, corresponds to the creation rate of 
electrons and holes. 
Originally, the creation rate for Dirac particles 
was calculated by Schwinger with the proper regularization 
method\cite{Schwinger1951}.
Below, we present a simpler derivation 
which can be extended to general band insulators.

The dynamics of the one-body model can be solved analytically, 
since we can cut the model into slices, each of which 
reduces to Landau-Zener's two band model\cite{Landau,Zener}.
Namely, if we decompose the $k$ vector
as $(\Vect{k}_\perp,k_\parallel)$, 
where $\Vect{k}_\perp$ ($k_\parallel$) is the 
component perpendicular (parallel) 
to $\Vect{F}$, each slice for a given $\Vect{k}_\perp$ 
is a copy of Landau-Zener's model 
with a gap $\Delta_{\rm band}(\Vect{k})\equiv 2\sqrt{V^2+v^2k_\perp^2}$.
The Landau-Zener transition takes place around the level anti-crossing
on which $k_\parallel+\frac{2\pi}{L}\phi(\tau )$ moves across the 
Brillouin zone(BZ) 
in a time interval $\delta \tau=h/F$.  The process can be expressed 
as a transition,
\begin{eqnarray}
c_+^\dagger(\Vect{k})\to 
\sqrt{1-p(\Vect{k})}e^{-i\chi(\Vect{k})}c_
+^\dagger(\Vect{k})+\sqrt{p(\Vect{k})}c_-^\dagger(\Vect{k}),\nonumber\\
c_-^\dagger(\Vect{k})\to -\sqrt{p(\Vect{k})}c_+^\dagger(\Vect{k})+
\sqrt{1-p(\Vect{k})}e^{i\chi(\Vect{k})}c_-^\dagger(\Vect{k}).
\label{2Bogoliubov}
\end{eqnarray}
Here the tunneling probability for each $\Vect{k}$ is given by 
the Landau-Zener(LZ) formula\cite{Landau,Zener}, 
\begin{equation}
p(\Vect{k})=\exp\left[-\pi
\frac{(\Delta_{\rm band}(\Vect{k})/2)^2}{vF} \right]
\label{LZS1},
\end{equation}
while the phase $\chi(\Vect{k})=-\theta(\Vect{k})+\gamma(\Vect{k})$ 
consists of the trivial dynamical phase,
$
\hbar \theta(\Vect{k})=
\int_0^{\delta \tau}\ve_{+}(\Vect{k}+
\frac{2\pi}{L}\phi(s)\Vect{e}_\parallel) ds,
$
and the Stokes phase\cite{Zener,Kayanuma1993},
\begin{eqnarray}
\gamma(\Vect{k})=\frac{1}{2}\mbox{Im}\;\int_0^\infty 
ds\frac{e^{-i(\Delta_{\rm band}(\Vect{k})/2)^2 s}}{s}
\left[\cot (vFs)-\frac{1}{vFs}\right].\nonumber
\label{2StokesphaseDDimBand}
\end{eqnarray}
The Stokes phase, a non-adiabatic 
extension of Berry's geometric phase \cite{Berry1984}, 
depends not only on the topology of the path but also on the
field strength $F$ \cite{Kayanuma1997}.
In terms of the fermion operators 
the ground state is obtained by filling the lower band
 $|0;\phi\ket=\prod_{\Vect{k}}
c^\dagger_-(\Vect{k}-\frac{2\pi}{L}\phi\Vect{e}_{\parallel})|{\rm vac}\ket
$, 
where $|{\rm vac}\ket$ is the fermion vacuum satisfying
$c_{\pm}(\Vect{k})|{\rm vac}\ket =0$.
If we assume that excited charges are absorbed by electrodes \cite{electrode}
we obtain from eqs.(\ref{ggamplitude}), (\ref{2Bogoliubov})
\begin{eqnarray}
\mbox{Re}\;\mathcal{L}(F) &=&-F
\int_{\rm BZ} \frac{d\Vect{k}}{(2\pi)^{d}}\frac{\gamma
(\Vect{k})}{2\pi},\nonumber\\
\mbox{Im}\;\mathcal{L}(F) &=&-F\int_{\rm BZ} 
\frac{d\Vect{k}}{(2\pi)^{d}}\frac{1}{4\pi}\ln \left[ 1-p(\Vect{k})\right],
\label{2Schwinger2}
\end{eqnarray}
where the dynamical phase $\theta$ cancels the 
factor $e^{\ihbar\int_0^\tau E_0(\phi(s))ds}$ 
in eq.(\ref{ggamplitude}). 
Performing the  $\Vect{k}$ integral in eq.(\ref{2Schwinger2}) 
leads to the ground-state decay rate per volume for
a $d$-dimensional hyperbolic band,
\begin{eqnarray}
\Gamma(F)/L^d&=&
\frac{F}{(2\pi)^{d-1}h}\left(\frac{F}{v}\right)^{(d-1)/2}\nonumber\\
&&\times
\sum_{n=1}^\infty \frac{1}{n^{(d+1)/2}}
e^{-\pi n\frac{V^2}{vF}}
\left[
\mbox{erf}\left(\sqrt{\frac{nv\pi^3}{F}}\right)\right]^{d-1}.
\label{BandSchwingerFormula}
\end{eqnarray}
To include the spin degeneracy we multiply this by $(2s+1)$ 
for a spin $s$ case.

If we compare eqs. (\ref{2Schwinger2}), (\ref{BandSchwingerFormula}) 
with Heisenberg-Euler-Schwinger's results 
of the nonlinear responses of the QED vacuum in strong electric 
fields\cite{Heisenberg1936,Schwinger1951}, 
two significant differences can be noticed.  
One is the error function appearing in eq.(\ref{BandSchwingerFormula}), 
which is due to the lattice structure (with the $\Vect{k}$
integral restricted to the BZ).  In the strong-field limit ($F\to \infty$), the 
$F^{(d-1)/2}$ factor in eq.(\ref{BandSchwingerFormula}) is cancelled, 
and the tunneling rate approaches a {\it universal} function, 
$\Gamma(F)/L^d\to -\frac{F}{h}\ln
\left[1-\exp\left(-\pi F_{\rm th}^{\rm band}/F \right)\right]$, 
independent of $d$.  Another, quantitative difference appears in the
Zener's threshold voltage\cite{Muller1977};
$F_{\rm th}^{\rm band}=V^2/va$ ($a$: lattice constant) for the 
dielectric breakdown is many orders smaller than
the threshold for the QED instability 
$F^{\rm QED}=\frac{m_e^2c^3}{\hbar}\sim 10^{16}\;\mbox{V/cm}$.

%%%%%%%%%%%%%%%%%%%%%%%%%%%%%%%%%%%%%%%%%%%%%%%%%%
{\it Dielectric Breakdown of a 1D Mott Insulator ---}
%%%%%%%%%%%%%%%%%%%%%%%%%%%%%%%%%%%%%%%%%%%%%%%%%%
Having clarified the one-body case, we 
now discuss the effect of the strong electron correlation on 
a non-linear transport, 
i.e., the dielectric breakdown in the one-dimensional Mott insulator.  
We employ the half-filled Hubbard model in static electric fields.
In the time-independent gauge, the Hamiltonian is $H+F\hat{X}$ with 
$
H=-t\sum_{i\sigma}
\left(c^\dagger_{i+1\sigma}c_{i\sigma}+\mbox{H.c}\right)
+U\sum_jn_{j\up}n_{j\dw},
$
and $\hat{X}=\sum_jjn_j$ the position operator\cite{Resta1998}. 
When $F=0$ the ground state is a Mott insulator 
having a many-body energy gap so far as  
$U>0$\cite{Lieb:1968AM,Woynarovich1982all}. 

After a strong electric field is switched on at $\tau=0$, 
quantum tunneling begins to take place, 
first from the ground state to the lowest excited states, 
and then to higher ones. 
The tunneling to the lowest excited states, which 
dominates the short-time behavior, has been shown 
to be understood with the Landau-Zener formula 
(see eq.(\ref{LZMott})) \cite{Oka2003}, 
but it is the long-time behavior that should be relevant to the 
actual breakdown.  

The ground-state-to-ground-state transition amplitude is, in the 
time-independent gauge, 
\begin{equation}
\Xi(\tau)=\bra 0|e^{-\ihbar \tau\left(H+F\hat{X}\right)}|0\ket e^{\ihbar \tau E_0},
\label{eq:transitionamplitude2}
\end{equation}
where we denote the ground state of $H$ as $|0\ket$ 
and its energy as $E_0$. 
The transition amplitude is calculated here numerically
by first obtaining $|0\ket$ in the open boundary condition 
with the density-matrix renormalization group 
(DMRG) method, and 
then solving the time-dependent Schr\"odinger equation 
for the many-body system with the time-dependent 
DMRG \cite{Vidal2004,White2004}.  The ground-state
decay rate $\Gamma(F)/L$ in strong electric fields $F$ is then obtained 
from the asymptotic behavior of $\Xi(\tau)$.

%%%%%%%%%%%%%%%%%%%%%%%%%%%%%%%%%%%%%%%%%%%%%
\begin{figure}[t]
\centering 
\includegraphics[width= 7.2cm]{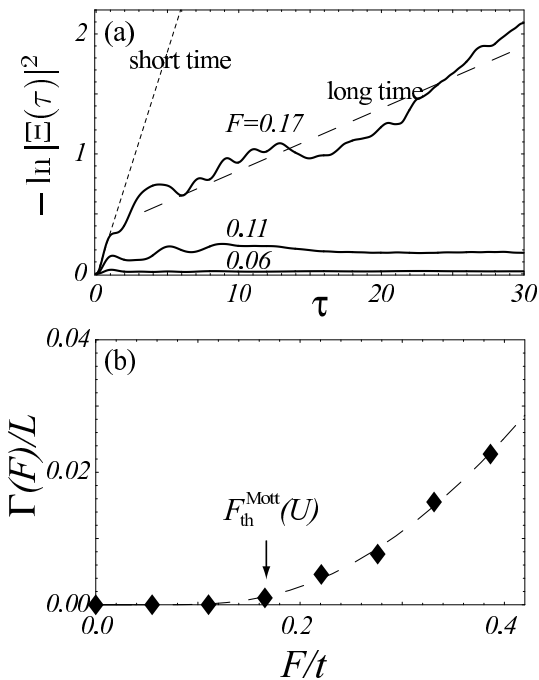}
\caption{
(a) The temporal evolution of the ground-state survival 
probability $|\Xi(\tau)|^2$ 
after the electric field $F$ is switched on at $\tau=0$ 
in the 1D half-filled Hubbard model with $U/t=3.5$, 
obtained with the time-dependent DMRG for $L=50$ 
with the size of the DMRG Hilbert space $m=150$
and the time step $d\tau=0.02$.
The dashed line represents $-\ln|\Xi(\tau)|^2=\Gamma(F)\tau+c$ 
for $F/t=0.17$.  
(b) The decay rate versus $F$
in the half-filled Hubbard model.
Dashed line is a fit to eq.(\ref{fittingfunction}),
where $F_{\rm th}^{\rm Mott}(U)$ is the 
threshold.
}
\label{ImSU35}
\end{figure}
%%%%%%%%%%%%%%%%%%%%%%%%%%%%%%%%%%%%%%%%%%%%
Figure \ref{ImSU35}(a) shows the temporal evolution of the 
ground-state survival probability $|\Xi(\tau)|^2$ 
for a system with $U/t=3.5$.  
We notice that, as time increases, 
the slope of $-\ln |\Xi(\tau)|^2$ decreases after an initial stage.  
The slope is proportional to the decay rate, 
so its decrease implies a suppression of the tunneling. 
We can regard this as evidence that charge excitations are initially 
produced due to the Zener tunneling, 
but that, as the population of the excitations 
grows, scattering among the excited states 
become important. This results in the ``pair annihilation" of carriers, 
which acts to suppress the tunneling rate. 
We have determined $\Gamma(F)$ 
from the long-time behavior with a fitting
$-\ln |\Xi(\tau)|^2=\Gamma(F)\tau+{\rm const}$. 

The decay rate per length $\Gamma(F)/L$
is plotted in  Fig.\ref{ImSU35}(b), 
where we have varied the system size ($L=30, 50$) 
to check the convergence.
$\Gamma(F)/L$ is seen to remain vanishingly small until 
the field strength exceeds a threshold. 
To characterize the threshold $F_{\rm th}(U)$ 
for the breakdown we can evoke the form obtained above 
for the one-body system (eq.(\ref{BandSchwingerFormula})), 
\begin{equation}
\Gamma(F)/L=-\frac{2F}{h}a(U)\ln
\left[1-\exp\left(-\pi \frac{F_{\rm th}^{\rm Mott}(U)}{F}\right)\right]
\label{fittingfunction}
\end{equation}
(with a factor of $2$ recovered for the spin degeneracy).  
The interest here is whether this holds when we replace 
$F_{\rm th}^{\rm band}$ with $F_{\rm th}^{\rm Mott}(U)$.
The factor $a(U)$ is a parameter 
representing the suppression of the quantum tunneling. 
The dashed line in Fig.\ref{ImSU35} (b) is the 
fitting for $U/t=3.5$, 
where we can see that the fitting, including the essentially 
singular form in $F$, is surprisingly good, 
given a small number of fitting parameters. 
The value of $a(U)$ turns out to be smaller than unity 
(taking between $0.77$ to $0.55$ as $U/t$ is increased from $2.5$ to $5.0$).

\begin{figure}[t]
\centering 
\includegraphics[width= 7.2cm]{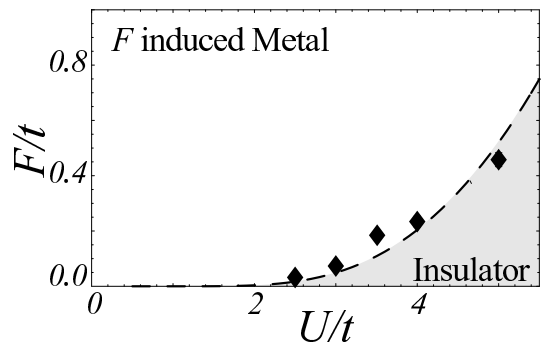}
\caption{The dielectric breakdown phase diagram 
on $(U,F)$ 
for the one-dimensional Hubbard model. 
The symbols are the threshold $F^{\rm Mott}_{\rm th}(U)$ 
obtained by fitting
the decay rate $\Gamma(F)/L$ to eq.(\ref{fittingfunction})), 
while the dashed line
 is the prediction $F=F_{\rm th}^{\rm LZ}(U)$
of the Landau-Zener formula eq.(\ref{LZMott}).
}
\label{Fth}
\end{figure}

In Fig.\ref{Fth} we plot the $U$ dependence of $F_{\rm th}^{\rm Mott}$.
The dashed line 
is the prediction of the Landau-Zener formula (see eq.(\ref{LZS1}))
\begin{equation}
F_{\rm th}^{\rm LZ}(U)=\frac{[\Delta_{\rm charge}(U)/2]^2}{v},
\label{LZMott}
\end{equation}
which was first applied to the dielectric breakdown 
of the half-filled Hubbard model in ref.\cite{Oka2003}.
For the size of the Mott (charge) gap we use the Bethe-ansatz result: 
$\Delta_{\rm charge}(U)=\frac{8t}{U}\int_1^\infty\frac{\sqrt{y^2-1}}
{\sinh(2\pi yt/U)}dy$\cite{Lieb:1968AM,Woynarovich1982all} with $v/t=2$.

%%%%%%%%%%%%%%%%%%%%%%%%%%%%%%%%%%%%%%%%%%%%%
%%%%%%%%%%%%%%%%%%%%%%%%%%%%%%%%%%%%%%%%%%%%

So the overall agreement between the threshold for the Hubbard model 
and eq.(\ref{LZMott}) is again confirmed, but to be more precise we 
note the following.  
Since the scattering among charge excitations 
suppresses tunneling, the 
threshold for the breakdown in interacting systems 
is expected to be larger than the Landau-Zener prediction (eq.(\ref{LZMott})). 
It was proposed in \cite{Oka2004a} that the wave function in 
electric fields just below the threshold 
is localized (on energy axis) around the ground state, 
which physically corresponds to a state where  
the production and annihilation of excited states are balanced 
due to the quantum interference. 
The $F/t=0.11$ result in Fig.\ref{ImSU35}(a) is typical of 
such states, where $-\ln|\Xi(\tau)|^2$ first increases
but soon saturates.  
When we further increase the field strength, 
the production exceeds the annihilation, 
which leads to an exponential decay of
the ground state with rate $\Gamma(F)/L$. 
The system is now metallic in the sense that
there are carriers that contribute to 
transport. 

As for the question of whether the dielectric breakdown in
the non-linear regime may be regarded as a true transition,
future studies should be needed.
The form (eq.(\ref{fittingfunction})) we used for fitting the tunneling rate
assumes that the transition
is a smooth crossover around $F\sim F_{\rm th}^{\rm Mott}(U)$.
However, the toy (quantum-walk) model calculation indicates that
the breakdown of many-body systems is
a localization-delocalization transition\cite{Oka2004a}.
This suggests that
the tunneling rate and other physical quantities
may become singular at the breakdown field, in which case
the $I$-$V$ characteristics exhibits a jump or a kink.
In our numerical calculation, while we do find an
indication of localization for $F/t=0.11$ in Fig.\ref{ImSU35}(a),
we observe no singularities in the tunneling rate.
So, future calculations will reveal the nature of the transition
around the breakdown. 

Finally, we comment on the relation of the 
effective Lagrangian approach 
with earlier approaches, especially the Berry's phase 
theory of polarization\cite{Resta1992,KingSmith1993,Resta1998,Resta1999,Nakamura2002}.  
There, the ground-state expectation value of the twist operator 
$e^{-i\frac{2\pi}{L}\hat{X}}$, which shifts the phase of electrons 
on site $j$ by $-\frac{2\pi}{L}j$ \cite{Nakamura2002}, plays a crucial role. 
It was revealed that the real part of a quantity 
\begin{equation}
w=\frac{-i}{2\pi}\ln\bra 0|e^{-i\frac{2\pi}{L}\hat{X}}|0\ket
\end{equation}
gives the electric polarization, 
$P_{\rm el}=-\mbox{Re}w$ \cite{Resta1998},
while its imaginary part gives a criterion for metal-insulator
transition, i.e.,
$D=4\pi\mbox{Im}w$ is finite in insulators 
and divergent in metals \cite{Resta1999}.
The effective action in the present work is regarded as 
a non-adiabatic (finite electric field) extension of $w$.
To give a more accurate argument, 
recall that the effective Lagrangian can 
be expressed as
$ \mathcal{L}(F)\sim \frac{-i\hbar}{\tau L}\ln\left(
\bra 0|e^{-\frac{i}{\hbar}\tau(H+F\hat{X})}
|0\ket e^{\frac{i}{\hbar}\tau E_0}\right)$ for $d=1$. 
Let us set $\tau=h/LF$ and consider the small $F$ limit.  
For insulators we can replace $H$ with the
ground-state energy $E_0$ to have 
\begin{equation}
\mathcal{L}(F)\sim wF
\end{equation}
in the linear response regime. Thus, the real part of 
Heisenberg-Euler's expression\cite{Heisenberg1936} for 
the non-linear polarization 
$P_{\rm HE}(F)=-\frac{\pa \mathcal{L}(F)}{\pa F}$ 
naturally reduces to the Berry's phase 
formula $P_{\rm el}$ in the $F\to 0$ limit (cf. eq.(\ref{2Schwinger2})).
Its imaginary part, which is related to 
the decay rate as $\mbox{Im}P_{\rm HE}(F)=-\frac{\hbar}{2}
\frac{\pa \Gamma(F)/L}{\pa F}$, reduces to $ -\frac{D}{4\pi}$
and gives the criterion for the transition, originally proposed
for the zero field case.

%%%%%%%%%%%%%%%%%%%%%%%%%%%%%%%%%%%%%%%%%%%%%%%%%%
%{\it Acknowledge:}
%%%%%%%%%%%%%%%%%%%%%%%%%%%%%%%%%%%%%%%%%%%%%%%%%%
We thank R. Arita for illuminating discussions.  
TO acknowledges M. Nakamura, S. Murakami, N. Nagaosa, and D. Cohen 
for helpful comments.
Part of this work was supported by JSPS.

After the completion of this work, we came to notice a recent
paper by Green and Sondhi\cite{Green2005}, where
non-linear transport near a superfluid/Mott 
transition is studied from the 
point of view of the Schwinger mechanism. 

%\bibliographystyle{prsty.bst}
%\bibliography{c:/Physics/documents2/ref.bib}

\begin{thebibliography}{10}
\bibitem{electrode}
This means that the  state vector after $n\delta \tau$ is
$\prod_{\Vect{k}}\left[\sqrt{1-p(\Vect{k})}e^{i\chi(\Vect{k})}
c_-^\dagger(\Vect{k})\right]^n|{\rm vac}\rangle$
since the $c_+^\dagger$ term disappears.
\end{thebibliography}
%\printindex

$^*$ Present address: CERC,
National Institute of Advanced Industrial Science and Technology,
Tsukuba Central 4, Tsukuba, Ibaraki, 305-8562 Japan.

\end{document}